\documentclass[%
 aip,
 amsmath,amssymb,
 reprint,%
]{revtex4-1}

\usepackage{graphicx}
\usepackage{dcolumn}
\usepackage{bm}

\usepackage[utf8]{inputenc}
\usepackage[T1]{fontenc}
\usepackage{mathptmx}
\usepackage{etoolbox}
\usepackage{xcolor}
\usepackage{textcomp}
\usepackage{color,soul}
\usepackage{caption}

\makeatletter
\def\@email#1#2{%
 \endgroup
 \patchcmd{\titleblock@produce}
  {\frontmatter@RRAPformat}
  {\frontmatter@RRAPformat{\produce@RRAP{*#1\href{mailto:#2}{#2}}}\frontmatter@RRAPformat}
  {}{}
}%
\makeatother

\begin{document}

\title{FSVPy: A Python-based Package for Fluorescent Streak Velocimetry (FSV)
}

\author{Han Lin}
\affiliation{ 
Department of Chemical and Biological Engineering, Northwestern University, Evanston, IL (USA)}

\author{Brendan C. Blackwell}
\affiliation{ 
Department of Physics and Astronomy, Northwestern University, Evanston, IL (USA)}

\author{Connor C. Call }
\affiliation{
Department of Chemical and Biological Engineering, Northwestern University, Evanston, IL (USA)
}%

\author{Shanliangzi Liu}
\affiliation{
Querrey Simpson Institute for Bioelectronics, Northwestern University, Evanston, IL (USA)
}

\author{Claire Liu}
\affiliation{
Querrey Simpson Institute for Bioelectronics, Northwestern University, Evanston, IL (USA)
}
\affiliation{
Department of Biomedical Engineering, Northwestern University, Evanston, IL (USA)
}

\author{Michelle M. Driscoll}%
\altaffiliation{Authors to whom correspondence should be addressed; electronic mail:
michelle.driscoll@northwestern.edu and jeffrey.richards@northwestern.edu}
\affiliation{ 
Department of Physics and Astronomy, Northwestern University, Evanston, IL (USA)
}%

\author{Jeffrey J. Richards}
\altaffiliation{Authors to whom correspondence should be addressed; electronic mail:
michelle.driscoll@northwestern.edu and jeffrey.richards@northwestern.edu}
\affiliation{ 
Department of Chemical and Biological Engineering, Northwestern University, Evanston, IL (USA)
}

\begin{abstract}

Predictive constitutive equations that connect easy-to-measure transport properties (e.g., viscosity and conductivity) with system performance variables (e.g., power consumption and efficiency) are needed to design advanced thermal and electrical systems. In this work, we explore the use of fluorescent particle-streak analysis to directly measure the local velocity field of a pressure-driven flow, introducing a new Python package (FSVPy) to perform the analysis. Fluorescent streak velocimetry (FSV) combines high-speed imaging with highly fluorescent particles to produce images that contain fluorescent streaks, whose length and intensity can be related to the local flow velocity. By capturing images throughout the sample volume, the three-dimensional velocity field can be quantified and reconstructed. We demonstrate this technique by characterizing the channel flow profiles of several non-Newtonian fluids: micellar Cetylpyridinium Chloride solution, Carbopol 940, and Polyethylene Glycol. We then explore more complex flows, where significant acceleration is created due to micro-scale features encountered within the flow. We demonstrate the ability of FSVPy to process streaks of various shapes, and use the variable intensity along the streak to extract position-specific velocity measurements from individual images. Thus, we demonstrate that FSVPy is a flexible tool that can be used to extract local velocimetry measurements from a wide variety of fluids and flow conditions.

\end{abstract}

\maketitle

\section{Introduction}

Quantifying the flow of complex fluids is critical to predicting their behavior in many technologies. Flow behavior is complicated by the presence of multiple phases, complex geometries, and non-Newtonian rheology, which leads to a wide variety of flow instabilities such as vorticity banding, shear banding, particle segregation, and slip. In these cases, the flow-velocity field must be directly visualized or computationally determined using simulations. Spatially resolved flow velocity visualization is a powerful tool to directly interrogate the flow behavior. Flow visualization techniques include direct {methods} which require tracer particles in the fluid, such as particle tracking velocimetry (PTV)~\cite{Hassan1991, Dracos1993, Maas1993, Hu2005, Masselon2008, Winer2014}, particle image velocimetry (PIV)~\cite{Kompenhans1987, Lourenco1988, Hain2007}, and streak velocimetry~\cite{Dimotakis1981}; and indirect/non-invasive techniques such as laser Doppler anemometry~\cite{Shapley2004}, nuclear magnetic resonance (NMR)~\cite{Callaghan2006}, and Doppler ultrasound methods~\cite{Poelma2017}.

Direct techniques including PTV and PIV have been extensively used to study the flow of complex fluids including rheometric flows{\cite{2006rheology}}. In PTV experiments the location of the particles is tracked between sequences of images and velocity is determined by the linear displacement of a particle using the known frame rate. In contrast, PIV determines displacement not by tracking individual particles, but by tracking the displacement of collections of particles between adjacent frames. The local velocity is then calculated by the displacement of that collection using the known frame rate. Both techniques have limitations that arise due to the requirement of tracking displacements between adjacent frames. For PTV this limitation manifests at high tracer particle densities{~\cite{Kahler2012}}, where the identity of the tracer particle can be lost in subsequent frames. While PIV performs better at high particle densities, it has a lower spatial resolution than PTV due to tracking collections of particles and produces bias errors in flows exhibiting large shear gradients where the configuration of particles within a collection can be significantly altered between adjacent frames{~\cite{Scharnowski2020}}.

While streak velocimetry is a direct method similar to PTV and PIV, a key difference is an increase in exposure time such that the trajectory of the particle is recorded on the image plane. In this way, the streak features including its length, shape, and changes in brightness can be directly related to the local flow field experienced by the tracer particle in a single frame{~\cite{Zhou2016, Chao2005, Salipante2020, Gbamele2000, Wang2018, Voss2012}}. While historically particle streak velocimetry{~\cite{Dimotakis1981}} has utilized scattered light for streak detection, recent examples have exploited phosphorescent particles to generate steaks. Fond et al. demonstrated that excitation of phosphorescent tracer particles could produce streaks whose length and decaying brightness encoded both the local flow velocity and the phosphorescent decay of the dye{~\cite{Fan2021}}. Recently, fluorescent tracer particles have been used to improve the resolution of PIV measurements{~\cite{Bergthorson2005, Kim2021, Walter2013, Qian2020, Santiago1998, Petrosky2015}}. For example, Petrosky et al. found that using fluorescent tracer particles reduced the laser flare near the wall of a flow channel enabling PIV in microchannels{~\cite{Petrosky2015}}. These findings suggest that fluorescent tracers have the potential to be utilized in streak applications, however, no systematic study has emerged that demonstrates the extraction of velocimetry information from fluorescent streaks in complex flow fields.

Here we describe a general framework to perform fluorescent streak velocimetry (FSV), which we define as the extraction of velocimetry data from streaks imaged in a single frame generated by the motion of fluorescent tracers in a flow field. This framework is integrated into a Python package (FSVPy) that uses existing feature-finding algorithms in scikit~\cite{scikit2011} to identify the orientation and location of streaks within an image. The velocimetry data within the streak is then analyzed using two frameworks that we validate in this work. The first is used for unidirectional flow where the streak intensity is fit to extract the streak width, which is proportional to the velocity of the tracer particle at that location. The second method is more general and determines the tracer velocity from the variation of fluorescent intensity along the streak contour, which we show is inversely proportional to the velocity of the tracer particle at that position. Both of these techniques return the quantitative local velocity of streaks detected within a single image frame. By acquiring many images over the entire flow volume, we show that the accumulated statistics on thousands of streaks can be analyzed to reconstruct the flow velocity field with micron-scale resolution in micro-fluidic and milli-fluidic flow channels with both Newtonian and non-Newtonian fluids.

\section{Materials and Methods} 
\subsection{Fluorescent Particle Synthesis}
In this work, all the chemicals were used as received without further purification. Rhodamine B, (3-Mercaptopropyl)trimethoxysilane (MPTPS), and ammonium hydroxide solution were all purchased from Sigma-Aldrich (MO, USA), Hydrogen peroxide 30 wt\% was purchased from Fisher (MA USA). Rhodamine-B doped silica particles were prepared based on Xin’s method\cite{Lu2013}. Briefly, 0.001 $g$ of Rhodamine B, 0.98 g of MPTPS, and 10 $mL$ of water were vigorously stirred in a round-bottom flask for 5 hours in the dark. Upon addition of the reagents, a turbid solution formed, as MPTPS is insoluble in water. The condensation reaction of the MPTPS drives the formation of silanol, which results in a slow transition to a transparent mixture during the reaction. At this stage, 0.5 $mL$ of ammonium hydroxide (1 wt\% aqueous solution) was rapidly dropped into the flask to trigger the condensation reactions of silanol, and the whole mixture was allowed to stand still for another 4 hours. 
After reaction completion, a pink sediment formed that settled to the bottom of the container. The sediment was dispersed in 3.4 $g$ of hydrogen peroxide and 3.4 $g$ of ethanol and stirred for 2 hours at room temperature. The final products were centrifuged and washed by DI water 5 times to remove the leftover dye and unreacted silane. Finally, all the solid sediments were collected and re-dispersed in water for storage. The morphology and size of the synthesized tracer particles was characterized by scanning electron microscope (SEM, JEOL JSM-7900FLV) after sputter coating with Au/Pt (EPIC SPF Desk IV). The FT-IR spectrum was characterized by a Nexus 870 spectrometer. Elemental composition was analyzed by X-ray photoelectron spectroscopy (XPS, Thermo Scientific ESCALAB 250Xi). The size of the oxidized particles was found to be 1.22 $\pm$ $0.16 \:\mu m$ by manually measuring 240 particles, as shown in Fig.\ S1. To confirm that oxidation was successful, we compared both XPS and FTIR results of the particles before and after oxidization, as shown in Fig.\ S2. 

\subsection{Materials}

For the glycerol data presented, pure glycerol from Fischer Scientific was diluted to an 85 $wt\%$ solution with milli-Q water. This is known to exhibit Newtonian behavior. We also tested a 12 $wt\%$ solution of cetylpyridinium chloride, sodium salicylate, and 0.5 M NaCl which we abbreviate as CPCl-Sal. At room temperature, this solution is known to form worm-like micelles that show strong shear-thinning and shear banding\cite{Hu2005,makhloufi1995rheo,Decruppe1995,CAPPELAERE1995353}. The rheology of this solution shows strong shear thinning and a power-law index close to -1, consistent with the presence of worm-like micelles. For the Carbopol data, Carbopol 940 from Sigma-Aldrich was dissolved in milli-Q water at a concentration of 0.1 $wt\%$ and neutralized to a pH of 7 with a 1M NaOH solution. This fluid is known to exhibit yield-stress behavior with minimal thixotropy. The 8 wt\% polyethylene glycol (PEG, 600 kDa, Sigma) solution was prepared by directly mixing with DI water. To dissolve all the PEG powder, the mixture was sonicated for 12 hours at room temperature. The prepared PEG solution was degassed in the vacuum oven before rheological measurement.

\subsection{Rheological Characterization}

Steady state flow curves for each material in the prior section were measured using a stress-controlled rheometer (DHR-2, TA Instruments) with a cone-plate geometry (40 $mm$ diameter and 1 degree cone angle) and a Peltier system at 25 $^{\circ}C$, and can be found in SI, Fig. S3b. See Fig. S3 of Supplemental Information (SI) for steady-shear curves. Solutions at varying concentrations of PEG, CPCl, and Carbopol were tested. The measurements were acquired using a flow sweep with shear rates logarithmically spaced from 0.1 – 1000 $1/s$. The measurement time at each shear rate was 60 s, which was found to be sufficient to ensure that the stress reached a steady state. For the samples containing PEG, we fit the resultant flow curves to the Carreau model, $\eta=\eta_{\infty}+\eta_{0}(1+a\dot{\gamma})^{-n}$.

\subsection{Fabrication of the Micro-channel}

\subsection{Imaging}
For the data shown in Fig.\ \ref{velocity_plot} and \ref{multi_mat}, the fluorescent colloids were dispersed in water at a concentration of approximately 200 ppm. Flow was induced via a 1 cm diameter glass syringe placed in a syringe pump, flowing through a 1 $mm$ $\times$ 1 $mm$ capillary that was placed on an Olympus IX83 inverted fluorescent microscope. Images were captured using a Hamamatsu ORCA-flash 4.0 camera, with a 20X objective and a doubler to achieve 40X magnification. The sample was illuminated with LED-generated 555 $nm$ wavelength light for an exposure time of 50 ms. Images were taken at intervals of 20 \textmu m in microscope objective height, with a correction to true channel depth applied during data processing in order to account for variable indices of refraction in the different liquids. Images were acquired at a frame rate of 2.3 fps. For the data shown in Fig.\ \ref{curved_streak}, \ref{velocity_extraction}, and \ref{expansion}, the colloids were dispersed in water at approximately 50 $ppm$ and flowed through PDMS channels, captured at frame rates from 4 to 10 fps with exposure times varying from 50 to 200 $ms$. All other methodology (pump, microscope, camera, illumination) was unchanged.
\subsection{Finite Element Simulations}
Three dimensional flow fields were simulated using OpenFoam\cite{OpenFoam}.
The channel was constructed using the dimensions of the channel and a mesh containing 1,600 nodes per cross section was constructed. SimpleFoam, a steady-state, incompressible fluid solver, was used with no-slip boundary conditions at all channel walls and a constant volumetric flow rate at the channel inlet. A zero-gradient boundary condition was applied to the channel outlet. Fully developed flow was observed within a few mm of the inlet and the cross-sectional velocity profile was taken downstream of the entrance region. 

\section{Results and Discussion}

FSV requires fluorescent tracers with robust fluorescence and slow photobleaching. In order to guarantee this, we modified an existing technique\cite{Lu2013} to produce hydrophilic Rhodamine B doped silica particles that exhibit a robust and long-lived fluorescent signal under the illumination conditions reported in this work. The synthesis and characterizations of tracer particles are summarized in the SI. To generate streaks, we illuminate a region of the channel with 555 nm light and increase the exposure time of the image acquisition until streaks are observed in the image plane.  Figure 1 illustrates typical streaks observed in unidirectional flow (Fig. \ref{streak_finding}a) and in a microfluidic chip with pillars (Fig. \ref{streak_finding}b). The raw images were bandpass filtered and the streaks were identified using find\_contours, a contour detection algorithm in skimage  that uses the marching squares algorithm to distinguish pixels as part of a feature~\cite{marching}. The corresponding filtered images and identified streaks are highlighted by their bounding boxes and numbered in Fig. \ref{streak_finding}c,d respectively. Each feature identified returns its bounding box, area, perimeter, angle, contour, and centroid, which can then be used for subsequent analysis as shown in Fig. \ref{streak_finding}e,f. This algorithm can robustly identify streaks regardless of their shape.

\subsection{Unidirectional Flow: Fitting streak length to quantify the velocity}

In order to demonstrate quantitative velocity detection in unidirectional flow, we acquire streak images of fluorescent tracer particles suspended in an 85\% glycerol solution in a flow channel with a square cross-section, 1 $mm$ $\times$ 1 $mm$, at a constant flow rate (2 $mL/hr$) controlled by a syringe pump. We image the flow channel with a 50 $ms$ exposure time. The streaks detected as described above are rotated based on the value of the contour angle $\alpha$, which is returned from the feature detection.  After rotation, the orientation of the streak aligns with the direction of the flow as shown in Fig.\ \ref{streak_plots}a. In unidirectional flow, we define the x-direction as the flow direction, the y-direction as perpendicular to the flow direction within the imaging plane, and the z-direction as normal to the imaging plane. We then averaged the fluorescent signal along the y- and x-directions from the centers of the bounding box with a 5-pixel bin as indicated by the red and blue boxes in Fig. \ref{streak_plots}a. The mean fluorescent intensity $I_{x0}(y)$ was fit to a Gaussian, 
\begin{equation}
    I_{x0}(y)=b+Ae^{-(y-y_0)^2/2h^2},
\end{equation} 
to determine the streak width, where $A$ is the amplitude, $b$ is the offset, $y_0$ is the pixel position of the streak center, and $h$ is the streak height. For the streak shown in Fig.\ \ref{streak_plots}a, the height was found $h=6.3$ pixels. Since the pixel width at this magnification is 0.1625  $\:\mu m$, the height of streak is $1.02 \:\mu m$ after calculation, which is consistent with the size of the particles determined by SEM. The mean fluorescent intensity along the y-direction $I_{y0}(x)$ was fit to a Gaussian function convoluted with a step function to determine the streak length $L$
\begin{equation}
    I_{y0}(x)=a+bx+(1-mx)\left[\mbox{erf}\left(\tfrac{x-x_0}{s}\right)-\mbox{erf}\left(\tfrac{x-x_0-L}{s}\right)\right],
\end{equation}
where $L$ is the streak length, $s$ is the width of distribution, $m$ is a slope of the intensity function, $b$ and $A$ defined as before.

This fitting procedure is repeated for each feature detected within a series of images acquired while scanning the entire channel cross-section along the y- and z-directions. All features were analyzed as above and the results for the streak heights and widths are associated with the properties determined from the find\_contours function as a dataframe. Within this population, we observe some anomalies including overlapping streaks, bubbles, and image artifacts. These can be removed from the population by filtering the total feature population based on the streak properties. We found that the streak height and angle provided the most effective filtering criteria, as the height of the streak should be narrowly distributed to reflect the distribution of particle size and the angle of the streak should be narrowly distributed to reflect the unidirectional nature of the flow. For the streak height, we fit the heights to a sum of three Gaussian functions and retained streaks that fell within 1 standard deviation of the most prominent peak height (Fig.\ \ref{streak_plots}b). We found that the mean of this filtered population agreed well with the particle radii. This filter eliminates streaks which are formed by overlapping two or more streaks and streaks that are out of focus. For the filter applied to the streak angle, a single Gaussian was fit to the observed distribution and we retained streaks that fell within 5 sigma of the most probably streak angle as shown in Fig.\  \ref{streak_plots}c. This works because all of the streaks should project along the flow direction, as our flow is unidirectional. The result of these two filters reduces the starting population of 4492 streaks to 2625 streaks and efficiently removes most anomalies.

Using this filtered population of streaks, we can reconstruct the velocity profile within the channel cross-section by averaging the velocity at each y-position and z-position as shown in Fig.\ \ref{velocity_plot}a, where the color corresponds to the measured velocity. The ability to determine a meaningful average of velocity depends on the choice of spatial resolution, the total number of frames acquired, and the number of streaks detected within each frame. We provide a simple geometric argument in the supplemental information that estimates the number of streaks per image for a given a spatial resolution, local velocity, and exposure time.  To verify our results, we simulate the fluid flow using OpenFoam\cite{OpenFoam}. The 2-dimensional cross-section of the velocity is shown at the same position down the flow channel in Fig.\ \ref{velocity_plot}b. The color of the velocity at each mesh point is on the same scale as that of our measurements, demonstrating good quantitative agreement between the measurements and the simulations. Fig.\ \ref{velocity_plot}c shows a slice of the 2-dimensional velocity profile along the z-direction, and additional slices represent the distribution of the velocities observed at each z-position. Note that many images were acquired at each position within the channel (30 per z-position for the data shown) such that many streaks are included in the reported averaged quantities (approximately 100 per z-position on average for the data shown). The most probable velocity along the x-direction is given as the solid points in Fig.\ \ref{velocity_plot}c, the red lines are histograms of the observed velocity distribution at each z-position, and the solid black line is a slice of the velocities determined from the OpenFoam simulation. There is quantitative agreement between the experimental and simulated velocity curves with the only adjustable parameter being the z-offset of the bottom wall. This demonstrates that FSV can be utilized to accurately quantify the velocity flow field of fluorescent tracer particles.

To demonstrate the wider applicability of FSV method and its ability to measure local flow velocity for complex fluids, we examine the CPCl solution. We acquired a series of images along the z-direction using 20 $\mu$m steps and averaged the identified and filtered streaks using the previously defined framework in Fig. 3. The resulting velocity profiles are shown in Fig. 4a for three different flow rates. The strong non-Newtonian character of these fluids is reflected in the plug-like flow at all flowrates. From the velocity profiles, We calculated the mean velocity for the 1, 2, and 4 $mL/hr$ cases and found them to be 0.29, 0.57, and 0.98 $mm/s$ respectively, which is comparable to the theoretical values of 0.27, 0.53, and 1.06 $mm/s$ expected for plug-like flow. At 2 $mL/hr$ pumping rate, we also compare profile of the CPCl-Sal sample to a glycerol solution, a 8 $wt\%$ PEG solution, and a 0.1 $wt\%$ Carbopol solution - a yield stress fluid. The resulting flow profiles are shown in Fig. 4b. Again, we observe complex flow behavior for all the samples with quantitative differences arising due to their different rheology. The viscosity-shear rate curves of each material are summarized in Fig. S3. The maximum velocities are plotted against power-law index in Fig. 4c, showing the expected relation of decreasing maximum velocity with increased shear-thinning.

\subsection{curvystreaks: an algorithm for identifying generic streaks}

The algorithm described in part A is computationally efficient, and works well to identify rectangular streaks. However, requiring the streaks to be rectangular limits our method to being used to measure flows which are unidirectional. To overcome this challenge, we created an additional computation package, curvystreaks, which uses a different and more flexible algorithm to identify the streak centerline.  While we have sacrificed computational efficiency, curvystreaks substantially increases the flexibility of the FSV method. To demonstrate this algorithm, we created a synthetic dataset created from a hand-drawn bezier curve; the identified centerline of this curve is shown in  Fig.\ \ref{curved_streak}a. 

In order to quantify velocimetry from these more complicated streaks, we utilize skeletonization, a well-known image-processing algorithm for identifying the medial axis of a binarized image. The first step is to binarize our image, using the polygon contour determined from the find\_contour function as a mask for binarization. Next, we apply skimage's medial\_axis function to this mask to return the medial axis of our streak (the medial axis is a good approximation to the centerline); this is the blue curve which is plotted in Fig.\ \ref{curved_streak}a. While medial\_axis robustly returns the centerline of our streak, the points are \emph{not} returned in their order along the centerline, but are ordered by their y-coordinate. As we would like to measure the velocity \emph{along} the streak to probe gradients in the flow velocity, this issue is severe. Fig.\ \ref{curved_streak}b illustrates this problem: the centerline points are colored according to their order.  As seen in that figure, the points as-returned are not ordered along the centerline. To resolve this issue, we developed a sorting algorithm which takes as an input the medial axis points as found via skeletonization and returns an ordered list of points. We note that solving this problem is equivalent to solving the traveling salesman problem. In other words, we need to identify the shortest path amongst a set of points for which we do not know the starting or ending point.

To sort the points, we borrow techniques from graph theory. First, we create a graph and add every found point as a node. We then use scipy's implementation of KDTree to create a distance tree.  We query this tree for all pairs which are 1.5 pixels apart; this identifies adjacent points. We then use this distance tree to populate the edges of the graph.  Next, we need to identify the source and target nodes of the graph, i.e.\ the starting and ending points. To do so, we generate a list of candidate source/target nodes by identifying the nodes which have a degree less than two (i.e., that only have one neighbor). We must now (1) identify the correct source and target node and (2) identify the shortest path which connects them, as this will be a good approximation to the true medial axis of our streak. To accomplish this, we measure by brute force all shortest paths between all candidate source/target nodes and then select the \emph{longest} shortest path; this will be the path that runs from the two end points of our streak along the medial axis. Fig.\ \ref{curved_streak}c shows the result of applying our sorting algorithm to the  unordered points shown  in Fig.\ \ref{curved_streak}b. It is clear from the figure that the true order of the points (along the streak) has been correctly identified.

An example of this algorithm applied to experimental curved streaks and used to measure velocity is shown in Fig.\ \ref{velocity_extraction}.  Fig.\ \ref{velocity_extraction}a shows the process applied to streaks generated from flow through a regular post array, and Fig.\ \ref{velocity_extraction}b shows the algorithm used on streaks generated by flow through an array of posts with randomly chosen radii and center position. The left panel shows the contours that are identified by the code plotted on top of the raw image. Once the streaks are identified and the points properly ordered (via the method explained in Fig.\  \ref{curved_streak}), the intensity as a function of position along each streak is plotted in the second panel, calculated using a rolling average of 3 points. The brightness at any given point on the streak is inversely proportional to the velocity of the particle as it travelled through that point: the slower the particle was going, the longer time it resided at that point, and the more light it emitted. The average velocity along the streak is known from the total length of the contour and the exposure time of the image (just as with the rectangular streaks in part A), and the average intensity along the streak is simple to compute once the data is in this form. From the average velocity, $v_{avg}$, average intensity, $I_{avg}$, and local intensity, $I_{l}$, it is then possible to compute the local velocity, $v_{l}$:
\begin{equation}
v_{l} = v_{avg}  \left( \frac{I_{avg}}{I_{l}} \right)
\end{equation}
The resulting velocity is plotted in the third panel as a function of position along the contour. Finally, the fourth panel shows the velocity as a function of location in the flow, plotted over a snapshot of the flow geometry that was taken in bright-field illumination. Note that the velocity results match physical intuition: streak 1 in the regular post array is moving faster than streak 0, which is far closer to  a solid boundary; the acceleration of streak 0 matches the pressure gradient that is expected for impinging flow, with the particle accelerating as it passes the obstacle then decelerating as it approaches the next obstacle. Also note that in the irregular post array, two streaks from the same region of the flow with different local intensities correctly measure the same velocity, which shows that this method is robust to variations in luminosity of individual tracer particles. For clarity of exposition, the example images illustrate only two streaks undergoing analysis, but each image in this data set averaged 6-8 usable streaks per image. The number of streaks per frame (SI Sect B) is dependent upon (1) the seeding concentration of tracer particles, (2) the exposure time of the image, and (3) the image size. Experiment duration was not a limiting factor in this data collection, but the density of velocity information per image can be increased for tests that cannot run for as long. Too high of a particle density will result in streak overlap, and too long an exposure time will result in too many streaks that touch the edge (the entire streak must be visible in order to know the average velocity in Eqn.\ 3), but both parameters can be pushed beyond what is shown here. These plots illustrate the power of the FSV technique: a single image allows one to simultaneously measure the velocity field at many varying locations in the flow field.
 
To demonstrate the construction of a full flow field in a velocimetric measurement, an expansion channel is shown in Fig.\ \ref{expansion}. Fig.\ \ref{expansion}a shows a schematic of the channel, and the streaks collected from 3000 frames of video are co-plotted in the upper half of Fig.\ \ref{expansion}b to show the full and detailed velocity map. The lower half of Fig.\ \ref{expansion}b shows a randomly selected sampling of the top half, mirrored across the dotted line. Fig.\ \ref{expansion}c shows a plot of velocity averaged over all y as a function of x, using 50 pixel wide bins in x, with the width of the channel co-plotted to show the relation. The data show the expected pattern for incompressible steady flow of linearly decreasing velocity with linearly increasing channel cross-section. The random sampling shown in the lower half of Fig.\ \ref{expansion}b does show some small variation in measured velocity of streaks near each other that are expected to have the same velocity. The cause of this variation is particles that are slightly higher or lower in z, but still within imaging range. The effect is stronger in this data than in Fig. 2-4 because the velocity gradient in z is larger (this channel is 100 $\mu$m tall, whereas the channel for the rectangular streak experiments was 1mm tall), creating a larger range of velocities present in the band in z where particles are in focus. Sufficient averaging alleviates any issues caused by this variation as demonstrated here, but using more sophisticated image processing to measure z-position via streak characteristics stands as a potential development that would further enhance this technique.

\section{Conclusions}

In summary, we have developed a robust and computationally efficient Python package for performing fluorescent streak velocimetry. We demonstrated the use of this code in a broad range of flow scenarios, first attaining 3D velocity fields for a both Newtonian and non-Newtonian fluids in unidirectional flow, then measuring more complex flow fields. The advantages of FSV over other commonly used techniques such as PIV and PTV are shorter run time and efficient encoding of information. It matches the spatial resolution of PTV, while containing all the information necessary to measure velocity in each single image, thus requiring no computationally-expensive cross-correlation steps. FSVPy is robust to streaks of any shape, and does not have stringent image quality requirements. This technique can produce detailed, three-dimensional velocimetric measurements in complex flows of complex fluids.

\section{Supplementary Material}

See supplementary material for the synthesis and characterization of the tracer particles, formulae for estimating the number of frames required in unidirectional flow, the rheology data of the non-Newtonian fluids, and schematics of micro-channels.

\section{Acknowledgements}

The authors appreciate the help from Prof. John A. Rogers of Northwestern University on the fabrication of complex micro-channels. We also acknowledge helpful discussions regarding graph theory algorithms with Istvan Kovacs. Han Lin was partially funded by the Department of Energy (DOE) Basic Energy Science Program (DE-SC0022119).  Han Lin and Brendan C.\ Blackwell were partially funded by Leslie and Mac McQuown through the Center for Engineering Sustainability and Resilience at Northwestern. This work made use of the EPIC facility of Northwestern University’s NUANCE Center, which has received support from the SHyNE Resource (NSF ECCS-2025633), the IIN, and Northwestern's MRSEC program (NSF DMR-1720139). Codes and tutorials can be found at {\href{https://github.com/mmdriscoll/FSVPy}{https://github.com/mmdriscoll/FSVPy}}.

\newpage

\begin{figure*}
    \centering
    \includegraphics[width=\textwidth]{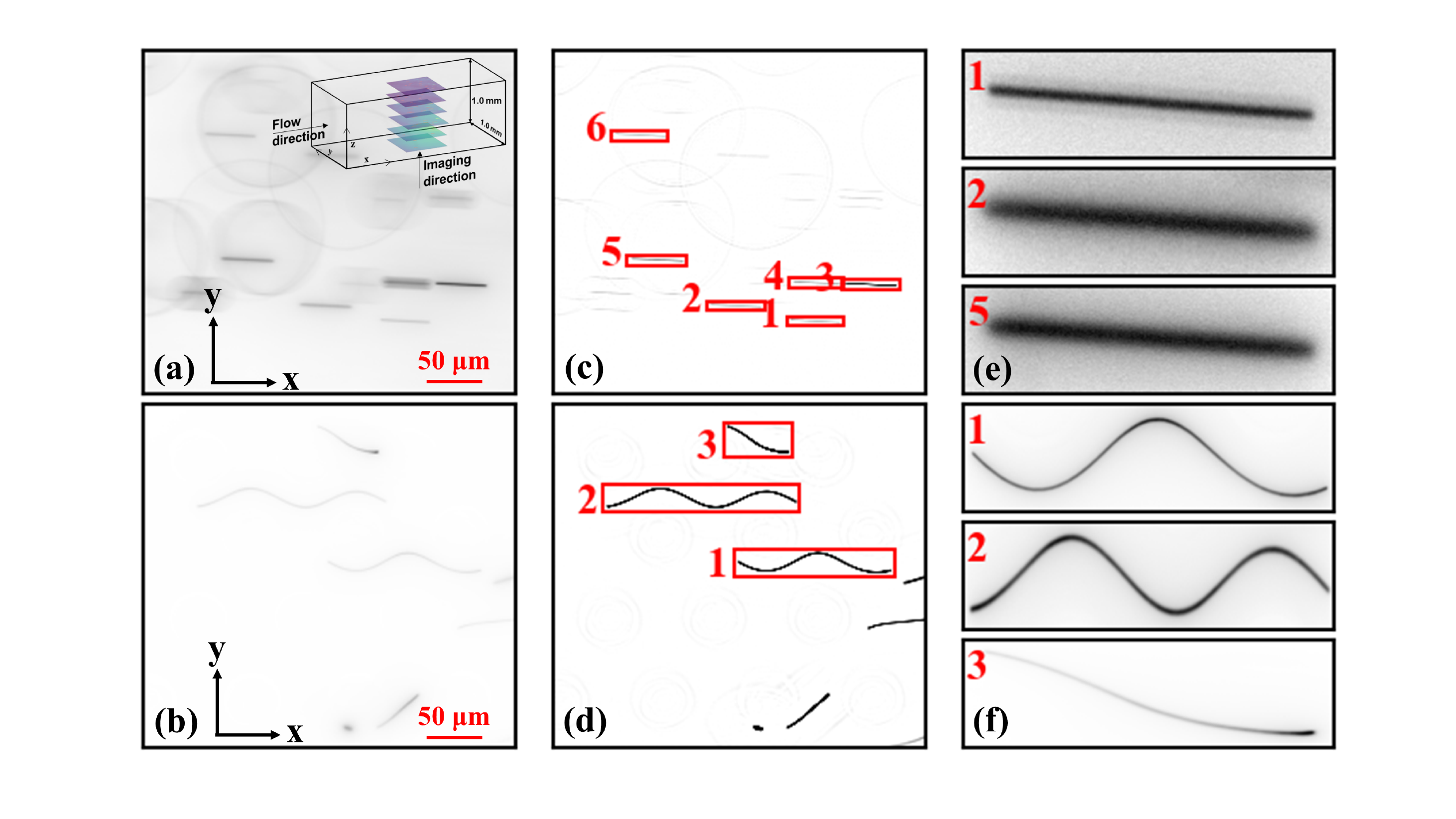}
    \caption{Illustration of the process from raw image to streak detection. (a) An unprocessed image with multiple streaks.  The inset depicts the schematic of the unidirectional flow channel. (b) The processed image after streak identification and filtering, qualified streaks are boxed with number. (c) Enlargements of individual streaks that were boxed in (b).   (d-f) Illustration of the same procedure for a complex flow that produces curved streaks.}
    \label{streak_finding}
\end{figure*}

\begin{figure*}
    \centering
    \includegraphics[width=\textwidth]{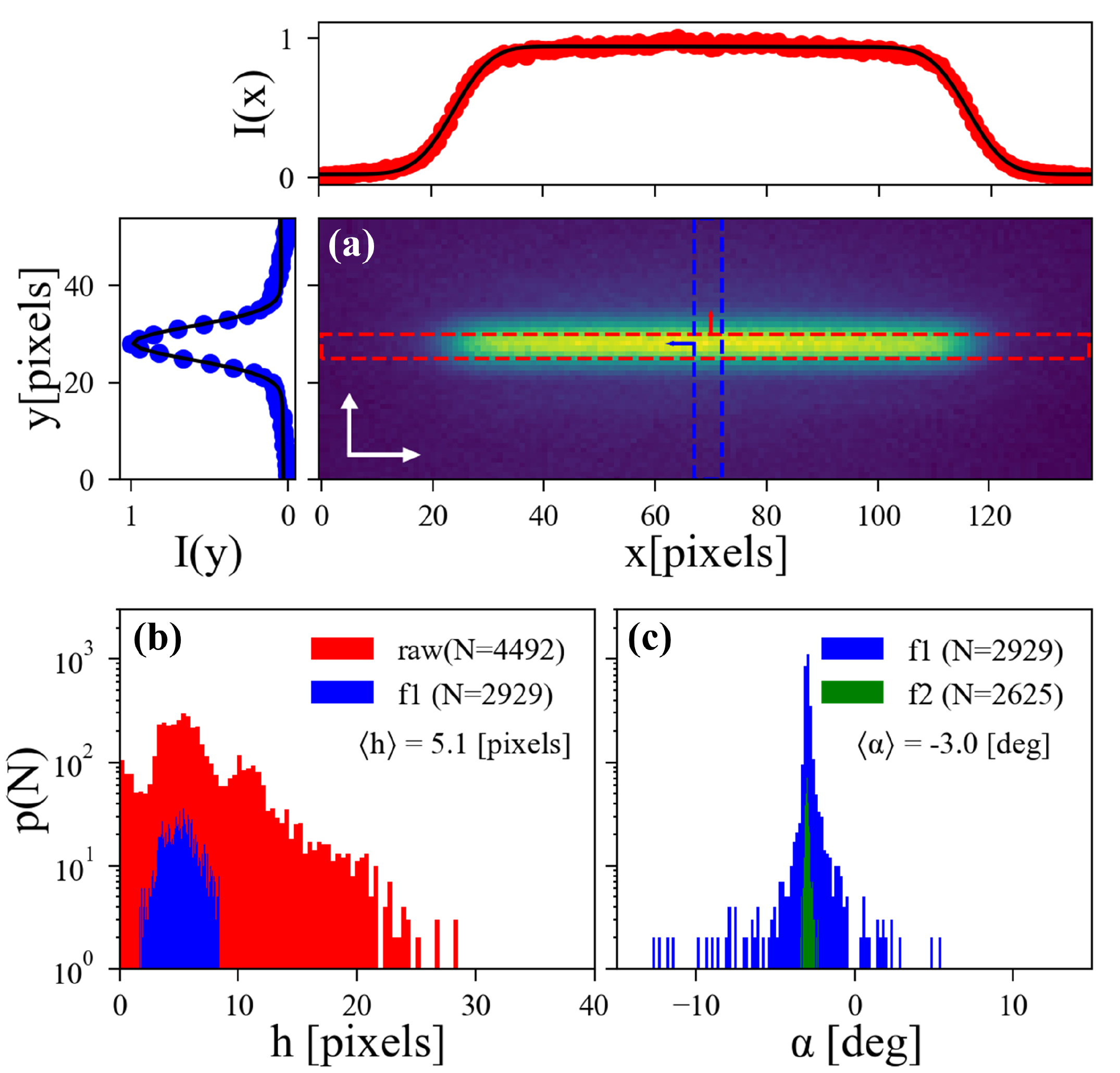}
    \caption{(a) An image of an individual streak after correcting for angle alignment. The red and blue dots are intensity along x and y axis, respectively. (b) Histogram of the height of all detected streaks (red) and the height of streaks used for data analysis (blue). (c) Histogram of the tilted angle of all detected streaks (blue) and the tilted angle of streaks used for data analysis (green).}
    \label{streak_plots}
\end{figure*}

\begin{figure*}
    \centering
    \includegraphics[width=\textwidth]{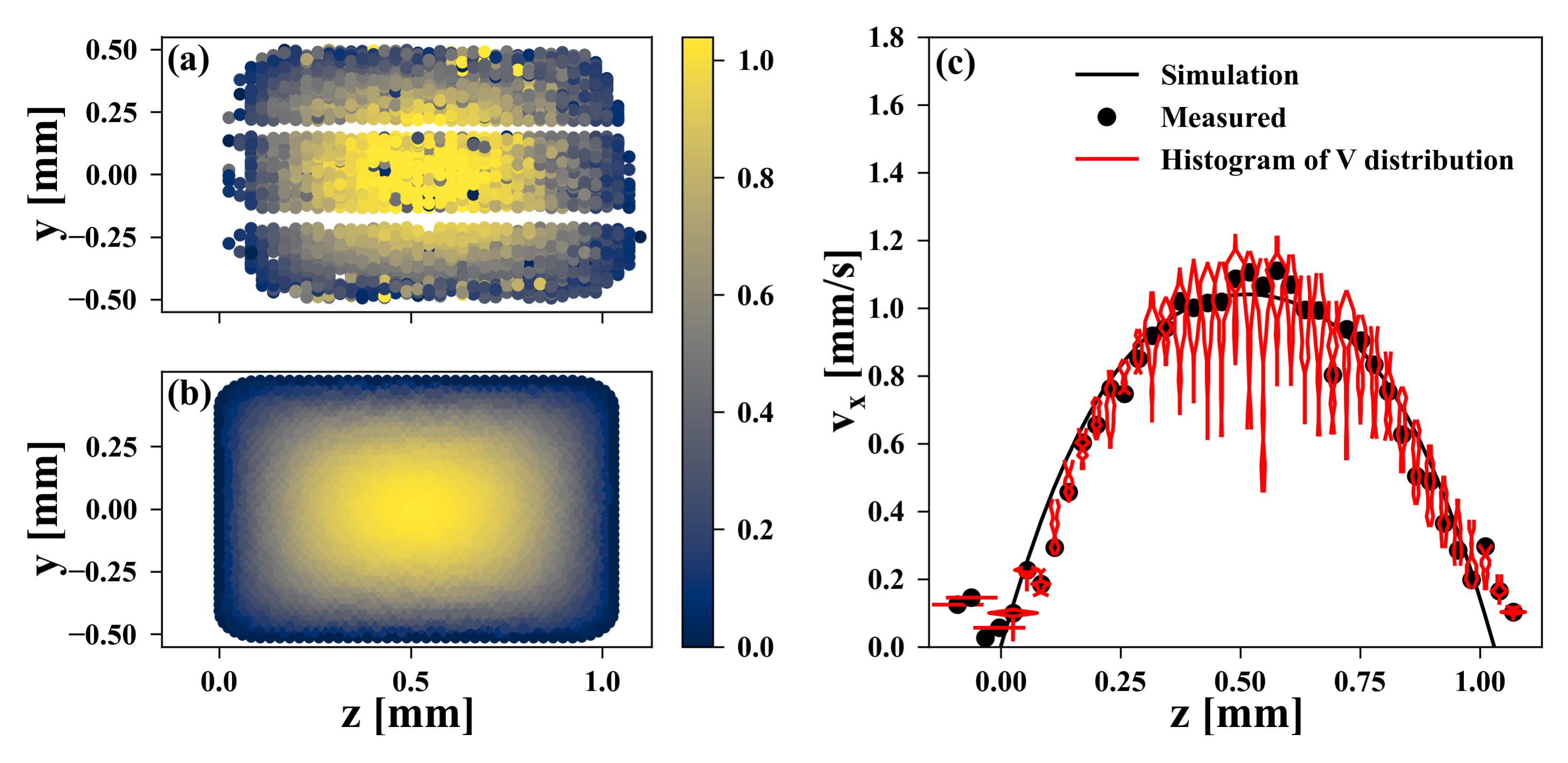}
    \caption{(a) The measured velocity map of the 85\% glycerol and (b) corresponding simulation velocity map of typical Newtonian fluid of this looking down channel. The unit of the colorbar is $mm/s$. (c) Velocity curve of 85\% glycerol at 2 $mL/hr$ along z-direction in the center of the channel. The red lines are histograms of the observed velocity distribution at each z-direction.}
    \label{velocity_plot}
\end{figure*}

\begin{figure*}
    \centering
    \includegraphics[width=\textwidth]{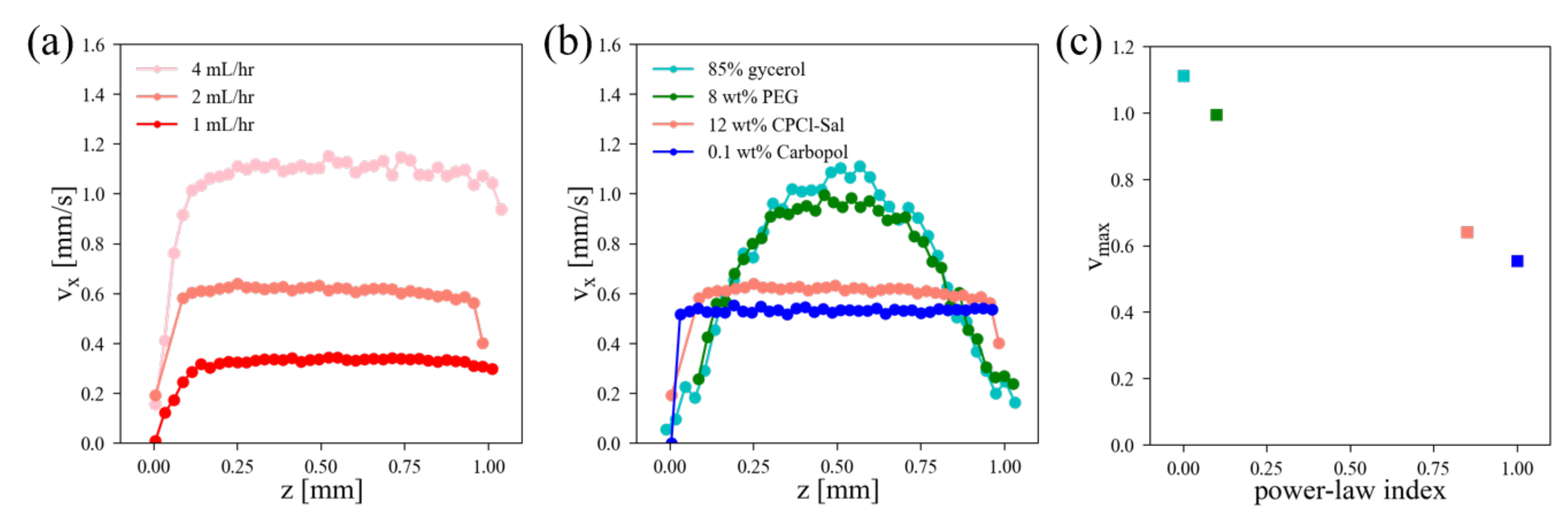}
    \caption{(a) Velocity of  12 wt\% CPCl-Sal measured along the z-direction at a flow rate of 1, 2, and 4 mL/hr. (b) Velocity of 12 wt\% CPCl-Sal, 0.1 wt\% Carbopol, 85\% glycerol, and 8 wt\% PEG measured along the z-direction at a flow rate of 2 mL/hr. (c) The maximum velocities observed in (b) as a function of the power-law index, which shows the strength of shear-thinning.}
    \label{multi_mat}
\end{figure*}

\begin{figure*}
    \centering
    \includegraphics[width=\textwidth]{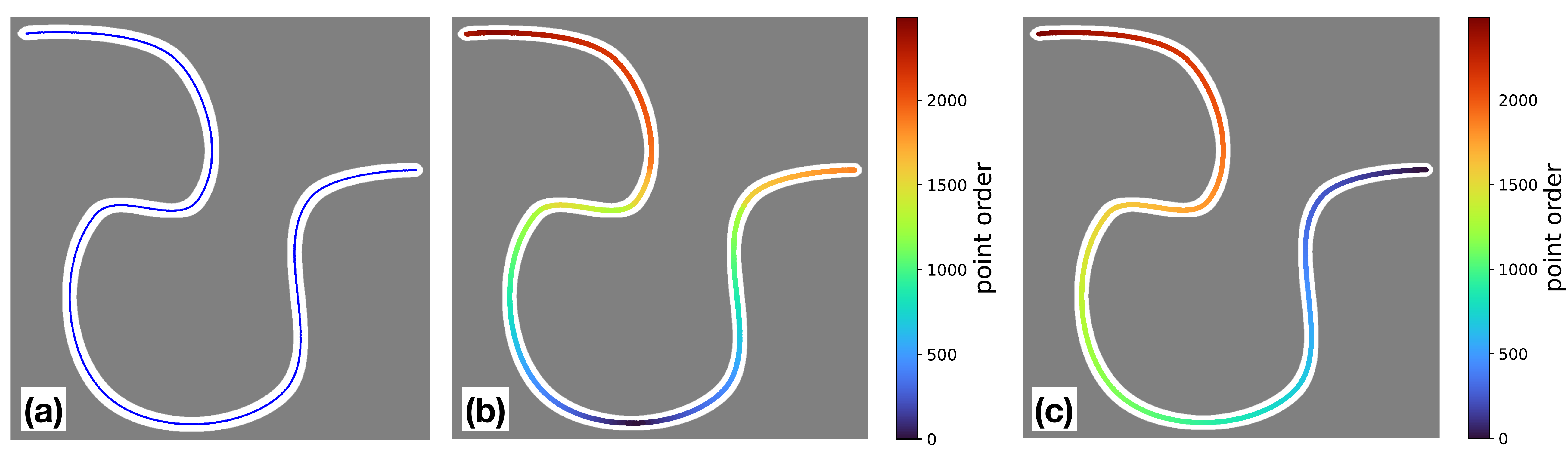}
    \caption{(a) All points of the centerline identified by the algorithm.  (b) The centerline points in (a) are colored according to their ordering.  As found, the points are  returned in the order of the y-coordinate. (c) After the sorting algorithm is applied, the set of points is ordered by their distance along the centerline.}
    \label{curved_streak}
\end{figure*}

\begin{figure*}
    \centering
    \includegraphics[width=\textwidth]{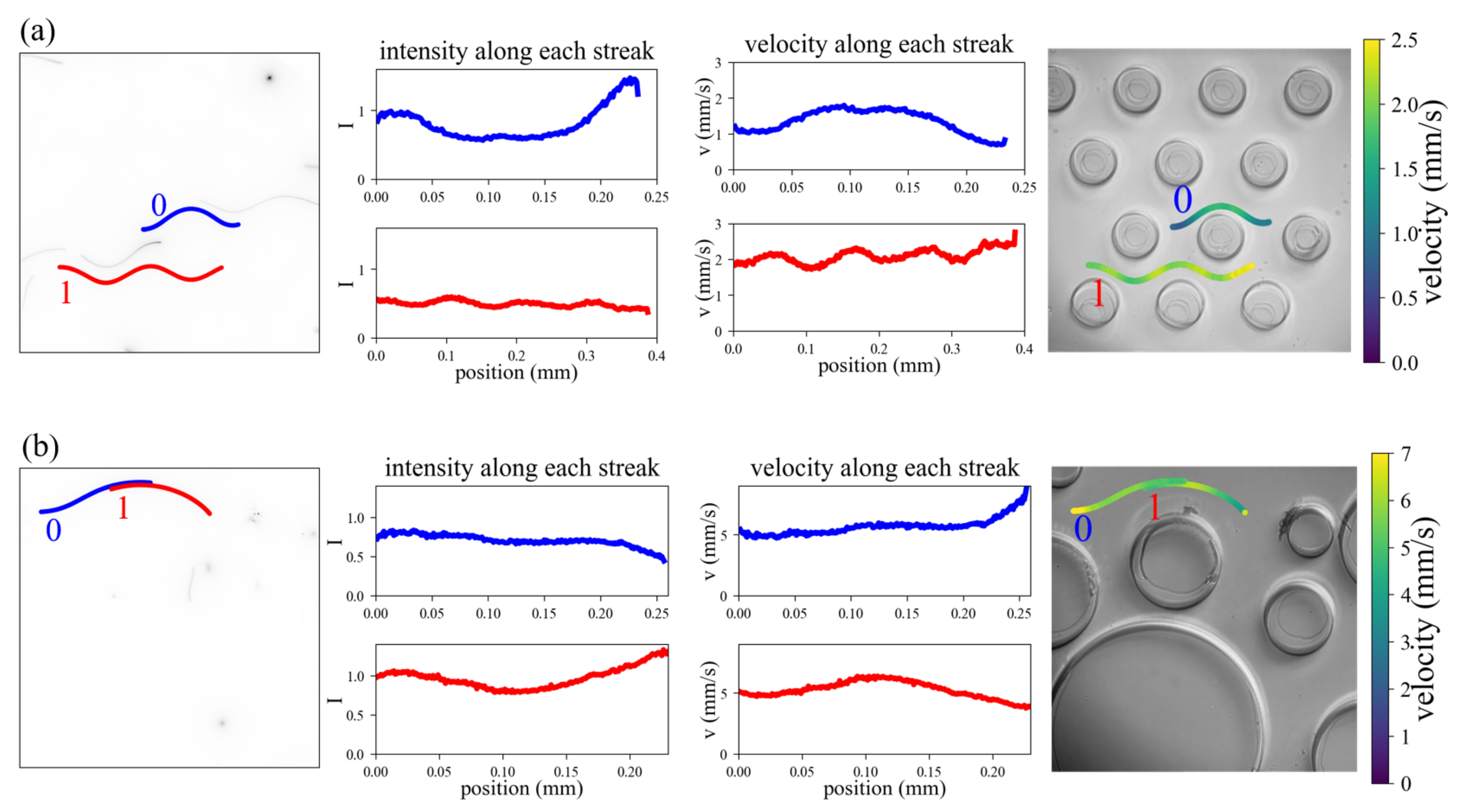}
    \caption{(a) An image of two example streaks in flow through an array of regular posts, with the Python-detected contours overlaid; intensity as a function of position along the streak for the example streaks; velocity as a function of position along the streak; velocity at each position on the streak, plotted on a snapshot of the flow geometry taken in bright field. (b) The same progression for two streaks in an array of randomly sized and positioned posts.}
    \label{velocity_extraction}
\end{figure*}

\begin{figure*}
    \centering
    \includegraphics[width=\textwidth]{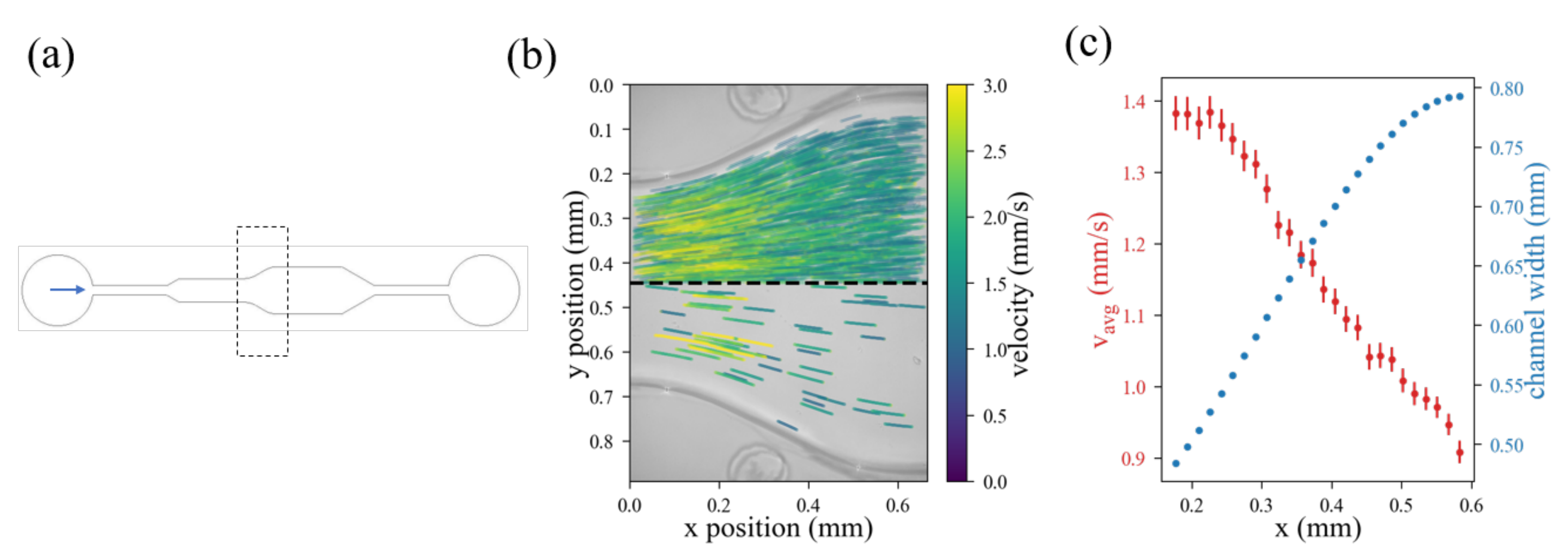}
    \caption{Velocity measurement of the flow in an expansion channel. (a) A drawing indicating the channel shape, with the dashed line indicating camera view, and the arrow indicating flow direction. (b) Plot of measured velocities. The upper half shows all streaks detected from 3000 frames of video, the lower half shows a random sampling of a few of the streaks plotted in the top half, mirrored across the dashed line. (c) The average velocity across all y values as a function of position in x.
    }
    \label{expansion}
\end{figure*}

\clearpage
\newpage

\section*{Supplemental Material}

\subsection{Tracer particle synthesis}
A two-step process was used to synthesize the hydrophilic Rhodamine B Silica Particles (RBSP). First, the silica particles with hydrophobic thiol functional groups were synthesized by following Xin's work\cite{Lu2013}. Second, to achieve better hydrophilic property, we applied an additional oxidation process using 30\% hydrogen peroxide. The hydrophobic thiol groups reacted with $\rm H_2O_2$ and formed hydrophilic sulfonic acids groups, as $\rm R-SH + 3H_2O_2 \rightarrow R-SO_3H + 3H_2O$. After the oxidization process, the well-dispersed RBSP in water was obtained, as the inset image shown in Fig. S1.

To study the morphology, SEM image was obtained on Au/Pt sputter coated RBSP, Fig. S1. The particle size distribution is generated based on the SEM images and about 240 particles were measured manually on Image J software. The mean and standard deviation of the diameter of RBSP is 1.2 $\pm$ 0.16 $\mu$m, which are good for detection by camera. To confirm the successful oxidized modification, the silica particles before and after oxidation were comparatively analyzed by XPS, Fig. S2. In high resolution S 2p measurement, a new peak was observed at about 168.4 eV\cite{yue1991} after oxidation, directly indicating the formation of $\rm SO_3^-$, as seen in Fig. S2a. FT-IR characterization was also used to confirm the successful oxidation on RBSP. Fig. S2 compares the FT-IR spectrum of RBSP before and after oxidation. First of all, the peaks of 2550 $\rm cm^{-1}$ were assigned to S-H adsorption\cite{xia2017}, that demonstrated the existence of thiol groups on the surface of particles, Fig. S2(b). Then, after oxidation, another obvious peak increase was at about 880 $\rm cm^{-1}$\cite{sax2018}, which represented the S-O stretching from newly formed $\rm SO_3^-$. Since S=O peak at about 1090 $\rm cm^{-1}$\cite{sax2018} overlapped with the typical silica peaks in range of 700-1300 $\rm cm^{-1}$, two curves were normalized by the peak intensity of Si-O-Si at 1030 $\rm cm^{-1}$\cite{siosi1,siosi2}, which was only from silica, as shown in Fig. S2c. The obvious increased intensity at 1090 $\rm cm^{-1}$\cite{sax2018} also demonstrated the formation of the new S=O bonds. What's more, the synthesized RBSP samples could remain stable after 3 month in water without dye leaking.\\

\subsection{Estimating the number of frames required in unidirectional flow}

For unidirectional flow, the number of frames required to obtain a desired statistical accuracy can be estimated as follows:

Assume:
\begin{enumerate}
    \item The imaging plane consists of $m \times n$ square pixels with the total image width $L$.
    \item The flow is steady and unidirectional such that $v_x=f(y)$ where the $x$ and $y$ directions represent the flow and flow-gradient directions on the imaging plane respectively.
    \item The image is exposed for a fixed time, $\tau$, and is acquired with an objective that produces a depth of field, $d$.
\end{enumerate}

We can then estimate the number of streaks per frame, $N_{st}(y)$ at a position $y$ with the desired spatial resolution, $\delta y$:

\begin{enumerate}
    \item The probability that a streak of length $l(y)=v_x(y)\tau$ will be wholly contained within the width of the image $L$
    \begin{equation}
        {p(\hat{l})=\frac{1-\hat{l}}{1+\hat{l}}\quad where\quad \hat{l}=\frac{l}{L}\le 1}
    \end{equation} 
 
    \item The probability that a streak will appear at a desired position $y$ with spatial resolution $\delta y$
    \begin{equation}
        {p(y)=\frac{\delta y}{n}\frac{m}{L}}
    \end{equation} 
 
    \item The average number of particles with volume, $V_p$, within the imaging plane, $\langle N \rangle$:
    \begin{equation}
        {\langle N \rangle=\frac{\phi}{V_{p}}\frac{L^2 n}{m} d}
    \end{equation}
  
    \item The number of whole streaks at a position $y$ in each frame is then given as:
    \begin{equation}
        {N_{st}(y)=p(\hat{l})p(y)\langle N \rangle}
    \end{equation}
\end{enumerate}

The frames to make a desired number of independent observations can then be calculated by selecting the desired spatial resolution, the exposure time, and the tracer particle density

\subsection{Rheology of solutions and micro-channel geometries}
We used a stress-controlled rheometer to measure the steady state flowcurve of these non-Newtonian fluids at 25 \textdegree C. A cone-plate geometry, which has a 40 mm diameter and 1 degree of cone, was used for all measurement. To ensure the obvious non-Newtonian behavior, the polyethylene glycol (PEG, 600 kDa) solutions are prepared with a weight\% of 0.5 - 8.0. A stable shear thinning behavior was observed for samples of 5.0 wt\% or more in the shear rate range of 0.1 - 1000 1/s. For 12 wt\% of CPCl-Sal, 0.1 wt\% and 0.25 wt\% of Carbopol samples, the measured results are within the shear rate range was 0.1- 100 1/s. We finally selected 8 wt\% of PEG, 12 wt\% of CPCl-Sal, and 0.1 wt\% of Carbopol samples for velocimetry measurements. The schematics of the fabricated micro-channels are shown in Fig. S4.


\begin{figure*}
    \centering
    \includegraphics[width=\textwidth]{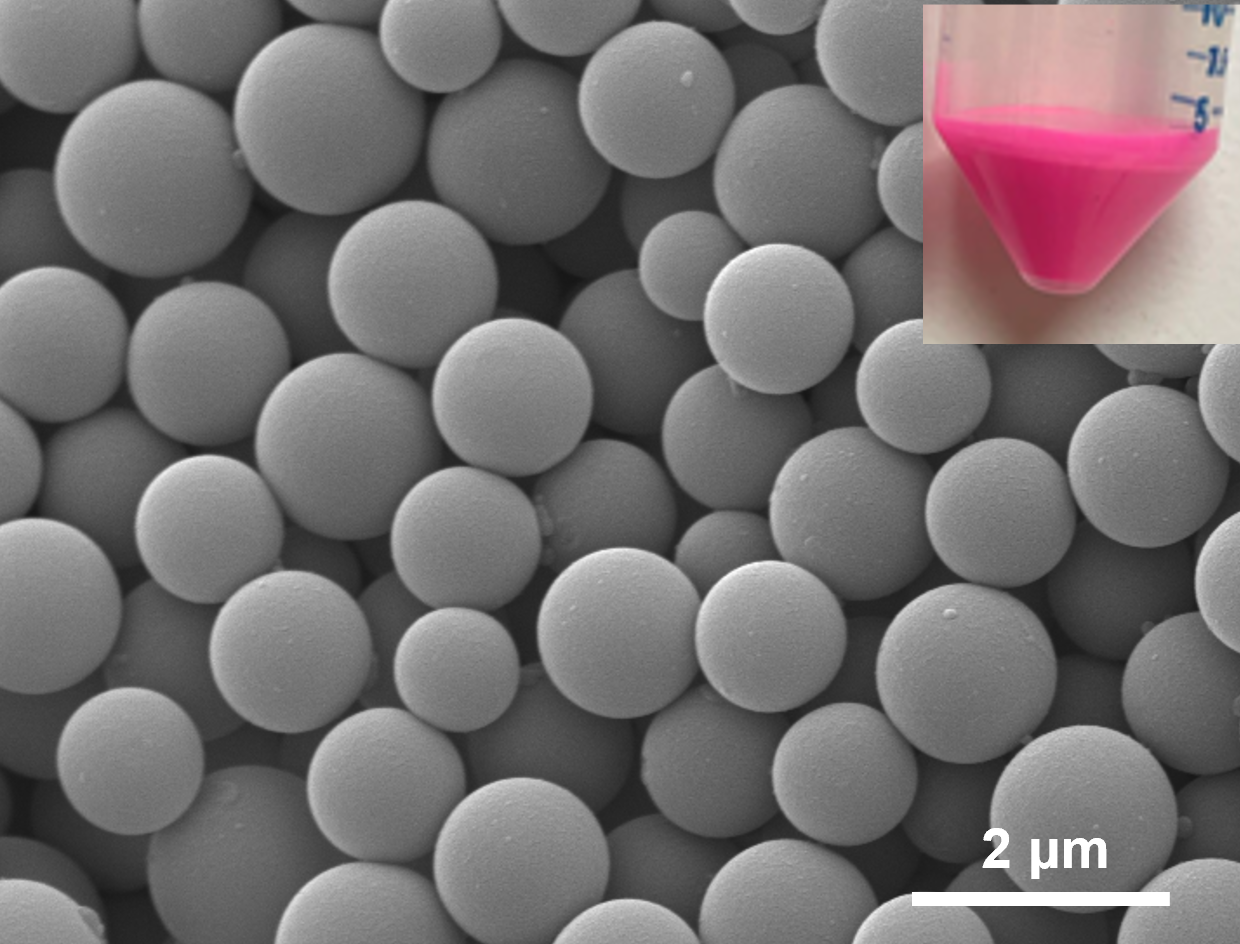}
    \caption*{FIG.\ S1: SEM images of synthesized RBSP. The inset digital image is suspension of RBSP in DI water.}
    \label{streak_finding}
\end{figure*}

\begin{figure*}
    \centering
    \includegraphics[width=\textwidth]{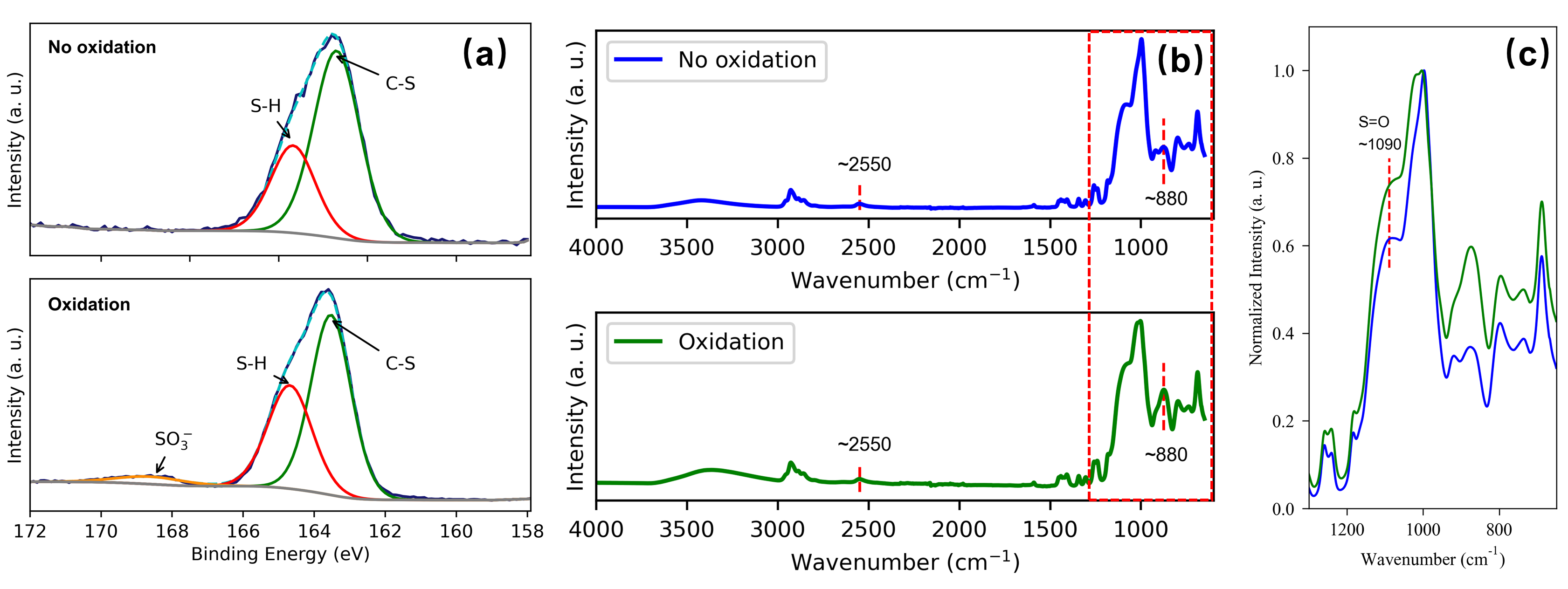}
    \caption*{FIG.\ S2: (a) The high resolution S 2p XPS results and (b) FR-IR results of RBSP before and after oxidation. (c) Normalized results of the box area of (b)}
    \label{streak_finding}
\end{figure*}

\begin{figure*}
    \centering
    \includegraphics[width=\textwidth]{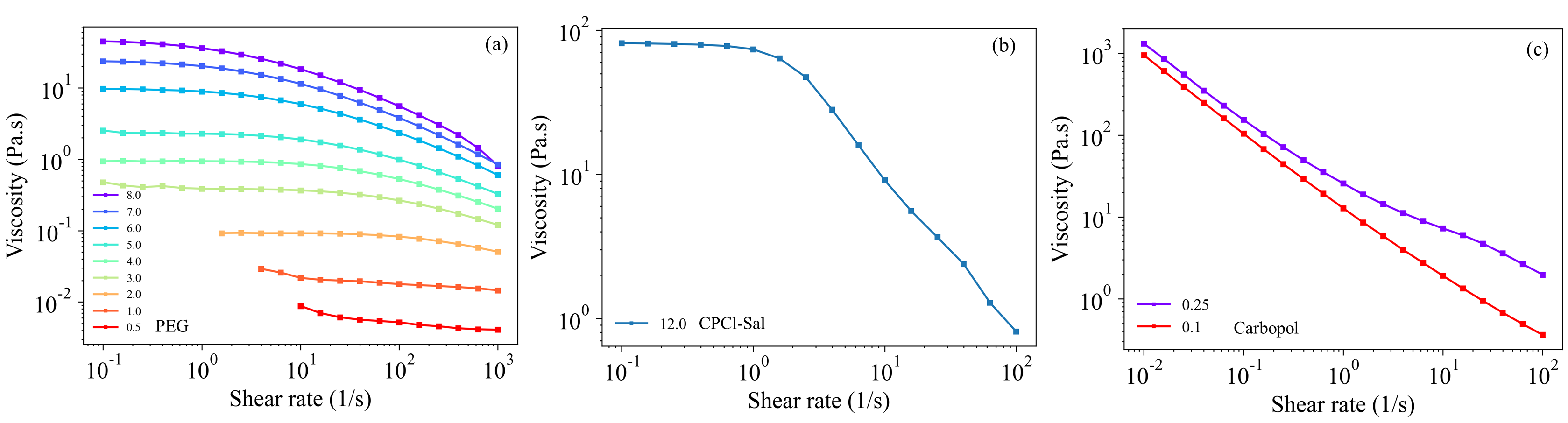}
    \caption*{FIG.\ S3: The shear rate vs.\ viscosity of (a) PEG (b) CPCl-Sal and (c) Carbopol. The numbers on the legends are the weight\% of each material. }
    \label{streak_finding}
\end{figure*}

\begin{figure*}
    \centering
    \includegraphics[width=\textwidth]{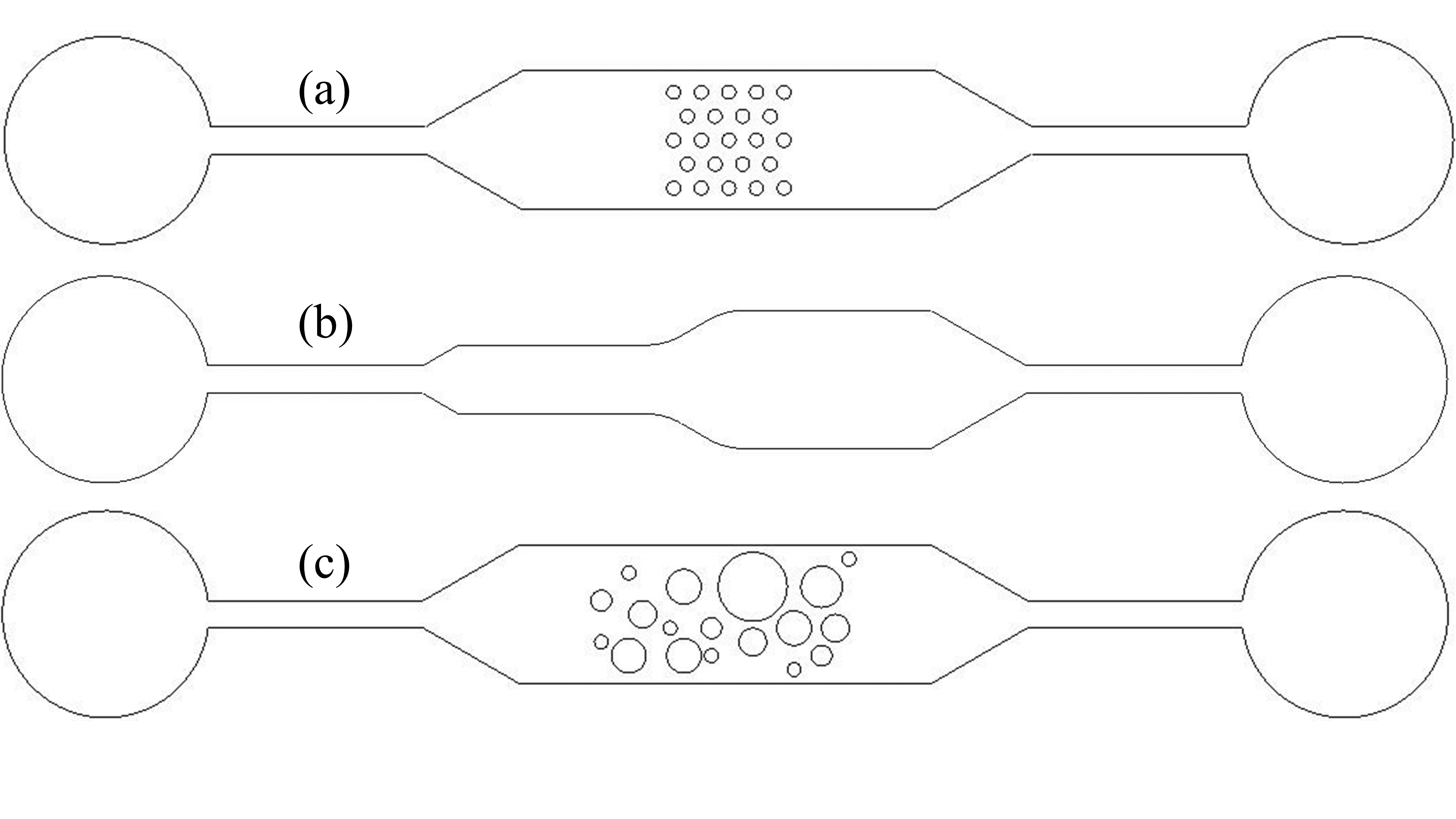}
    \caption*{FIG.\ S4: The geometries of the fabricated micro-channels. (a) The regular pillars micro-channel, (b) the expansion micro-channel, and (c) the irregular pillars micro-channel.}
    \label{streak_finding}
\end{figure*}

\bibliographystyle{unsrt}
\bibliography{VISER}


\end{document}